# Chiral Quasi-Bound States in the Continuum

Adam Overvig[1,2], Nanfang Yu[2*], and Andrea Alù[1,3*]

[1]Photonics Initiative, Advanced Science Research Center at the Graduate Center of the City University of New York

[2]Department of Applied Physics and Applied Mathematics, Columbia University

[3]Physics Program, Graduate Center, City University of New York

*Quasi-bound states in the continuum (q-BICs) are resonant states of suitably tailored nanostructures with long optical lifetimes controlled by symmetry-breaking perturbations. While in planarized ultrathin devices the resulting Fano resonance is limited to linear polarization, we show here that chiral perturbations extend q-BIC concepts to arbitrary elliptical polarizations. Using geometric phase engineering, we realize metasurfaces with ultrasharp Fano spectral features that can shape the impinging wavefront with near-unity efficiency, while at the same time precisely filtering their spectral content.*

Suitably designed photonic crystal slabs (PCSs) have been recently shown to support quasi-bound states in the continuum (q-BICs), opening new opportunities to enhance and control light-matter interactions. Q-BICs are modes whose radiative lifetime is controlled by a symmetry-lowering perturbation [1]; they would be ideally non-radiating states [2]-[4] due to symmetry-protection in the absence of perturbation, despite their momentum being compatible with coupling energy to the radiation continuum. When light with a polarization state matching the eigenpolarization of the q-BIC impinges on the structure, an ultrasharp Fano response arises [5], and the resonantly scattered light maintains the same polarization. This property, combined with strong in-plane Bragg scattering in high-contrast index systems [6], enables compact optical devices concentrating light in both space and time [7]-[11]. By perturbing every other unit cell in such systems, the Brillouin zone folds, enabling access to previously bound modes and providing additional design freedom to control q-BICs in real- and momentum-space [10]-[13]. Photonic crystals supporting q-BICs hence offer a highly versatile platform for biological sensing [14], planar optical modulators [15], notch filters [16] and nonlinear optics [17],[18].

Recently, the selection rules for q-BICs in planar photonic crystals have been classified for both mono-atomic and multi-atomic lattices, clarifying to which (if any) free-space polarization state a q-BIC may couple due to a chosen symmetry perturbation [13]. This result implies that in planar structures that preserve symmetry across a horizontal plane, the q-BIC polarization must necessarily be linear. By adiabatically varying the polarization angle of the supported q-BICs in the lateral direction and thereby introducing a spatial variation of the geometric phase, it is possible to tailor the impinging wavefront and realize metasurface functionalities, such as anomalous reflection and refraction for circularly polarized light emanating from the metasurface [19]. However, due to the planarized symmetry and the linear polarization constraint, the maximum achievable efficiency of wavefront shaping is $25\%$ [20].

Breaking this symmetry with the introduction of optical chirality may enable devices with strong circular dichroism [22]-[29]. Recent work has demonstrated metasurfaces that leverage this principle by varying the geometric phase associated with spin-preserving chiral structures in devices with asymmetric transmission profiles [30]-[33]. In parallel, the emerging field of twistronics has been exploring how anisotropic parallel layers stacked with a twist with respect to one another can exhibit unexpected electronic [34] and optical [35],[36] properties pertaining to the eigenstates of the composite system, opening new degrees of freedom in engineering wave-matter interactions.

Motivated by these explorations, in this Letter we explore two-interface systems where chiral perturbations produce q-BICs with novel optical properties. In particular, we study a pair of tightly stacked interfaces that scatter independently controlled linearly polarized waves, resulting in a composite metasurface supporting chiral q-BICs. While applying symmetry perturbations to an achiral photonic crystal necessarily yields a linear eigenpolarization, we demonstrate that chiral perturbations produce arbitrary elliptical eigenpolarizations of the supported q-BIC. The elliptical state is controlled by the orientation of the linearly polarized scattering at each interface, supporting a Fano resonance with full elliptical dichroism. Next, by tuning a geometric phase associated with the in-plane orientation of the chiral perturbations, while simultaneously detuning from the fully circular dichroic case, we show an arbitrarily sharp circularly polarized Fano resonance with tailored reflected amplitude and phase when illuminated by circularly polarized light. By adiabatically varying in space this geometric phase, we then demonstrate a metasurface supporting a generalized Fano resonance offering nearly transparent response at all frequencies other than an ultranarrow spectrum over which anomalous beam steering in reflection with near-unity diffraction efficiency is observed.

Consider a slab supporting a symmetry-protected BIC driven by two homogenized interfaces engineered to scatter the resulting q-BIC to the far field by breaking a relevant symmetry. Any system supporting a q-BIC in such a dielectric platform must have a non-negligible thickness (on the order of the effective wavelength), meaning that the minimal characterization of the composite metasurface (regardless of chirality) involves two interfaces separates by a non-negligible distance. We assume that the bottom and top interfaces scatter light to linear polarization states oriented at angles $\phi_1$ and $\phi_2$, respectively, where in the achiral case [Fig.1(a)], $\phi_1 = \phi_2$ due to symmetry, while in the chiral case [Fig.1(e)] $\phi_1$ and $\phi_2$ can vary at will. Such a metasurface may be implemented by properly chosen perturbations applied to an otherwise high-symmetry PCS [Figs. 1(b,f)].

We begin by studying the eigenpolarizations of an achiral PCS with square lattice periodic in the $x$ and $y$ directions and mirror-symmetric in the $z$ direction. Shown in Fig.1(b), the device sits on a substrate with refractive index $n_1 = 1.45$ and is composed of a thin film with refractive index $n_2 = 3.45$ with etched *X*-shaped inclusions filled with the superstrate material ($n_3 = n_1$). Each unit cell contains two *X*-shaped inclusions oriented 90° relative to one another and spaced with the periodicity of the square lattice, $a$. The resulting dimer supports modes that are bound in the unperturbed lattice but folded into the continuum by the dimerization [13]. To determine the resonant eigenpolarization, the structure may be decomposed into two perturbations, $V_1$ and $V_2$, applied to the unperturbed monatomic square lattice $H^0$ [Fig. 1(c)]. Selection rules [13],[40] predict that $V_1$ scatters the q-BIC to $y$-polarized light, while $V_2$ scatters it to $x$-polarized light. As depicted in the homogenized model [Fig.1(a)], a superposition of $V_1$ and $V_2$ cannot control $\phi_1$ and $\phi_2$ separately. Instead, their linear superposition results in the eigenpolarization $\phi_1 = \phi_2$, i.e., a real-valued superposition of $y$- and $x$-polarized light yielding a linear polarization angle that may be arbitrarily tuned by controlling the magnitude of $V_1$ and $V_2$ [19],[40]. As shown in Fig. 1(d), irrespective of $\phi_1$, incident circularly polarized light couples only half of the incident power to the resonance, regardless of its handedness (left-handed, LCP, or right-handed, RCP). This is expected, since the device is achiral and so cannot exhibit circular dichroism.

In order to overcome this constraint, we analyze the metasurface response when two perturbations $V_1$ and $V_2$ are applied to the two interfaces independently. This class of symmetry perturbations breaks the mirror symmetry about the horizontal plane and introduces a phase difference $\Delta\Phi$ between the coupling via each perturbation, enabling a chiral response. The resulting metasurface, shown in Fig. 1(f), consists of two layers of etched elliptical inclusions, which may be implemented with multi-step lithography and planarization [41],[42]. The composite perturbation, shown in Fig. 1(g), is characterized by the in-plane orientation angle $\alpha$ of the ellipses in the bottom layer ($V_1$) and the difference in orientation angles $\Delta\alpha$ between the bottom layer and the top layer ($V_2$). The selection rules prescribe coupling to light polarized at angles $\phi_1$ (following $\phi_1 \approx 2\alpha$) for $V_1$, and $\phi_2$ (following $\phi_2 \approx 2(\alpha + \Delta\alpha)$) for $V_2$ [40]. Both of these linearly polarized states scatter through the superstrate and substrate, interfering with direct optical pathways to produce a Fano resonance. Scattering at the internal interface (i.e., at the junction between the stacked ellipses) is negligible, as the two regions are different only by a small perturbation. This system is therefore well-described by the homogenized device depicted in Fig. 1(e), as confirmed by temporal coupled mode theory [40].

Importantly, the introduction of a phase difference implies that the superposition of light scattered to each direction is weighted by a phase-factor $e^{i\Delta\Phi}$, which accounts for the relative phase of light coupling through the substrate and superstrate. It is therefore determined by both the local background scattering process (the propagation phase through the slab) and the modal decay symmetry, which for small chirality is symmetric or anti-symmetric [37]. When the slab is a quarter-wavelength thick, i.e., $h = \lambda / 4n_{eff}$, where $h$ is the height of the slab and $n_{eff}$ is the effective slab index, the phase-factor is $e^{i\Delta\Phi} = (-1)^{n-1} i$, where $n$ is the number of lobes that the q-BIC has in the $z$ direction. In this case, the eigenpolarization is $\phi_1 \pm i\phi_2$, as shown in Fig. 1(g). If $\Delta\alpha$ is chosen such that $\phi_2 = \phi_1 \pm 90^{\circ}$, the eigenpolarization is RCP. Figure 1(h) shows the reflection for opposite handedness when $\alpha = 70^\circ$ and $\Delta\alpha = 50^\circ$, demonstrating full circular dichroism at the resonant wavelength: RCP light is fully resonantly reflected, while LCP light does not couple at all to the resonance. The bandwidth of this resonance is tunable by the magnitude of the perturbation $\delta = D_a - D$, as usual for q-BICs [1],[13], and it can be made extremely sharp for small $\delta$.

We now vary $\Delta\alpha$ from –90$^\circ$ to 90$^\circ$ and $\alpha$ from 0$^\circ$ to 90$^\circ$, and study the eigenpolarizations of the corresponding q-BICs by analyzing the Jones matrix $M_t$ of the two-interface system for the transmitted light. At the resonant frequency, the eigenpolarizations of the scattered waves are the eigenvectors of $M_t$: they are the polarization states that are unaffected by the scattering process, and the corresponding eigenvalues are the scattering amplitudes of these polarization states. Since the metasurfaces in Fig. 1 operate near a transmission maximum of the background scattering process, the non-resonant eigenpolarization corresponds to the eigenvector of $M_t$ with near-unity eigenvalue. The vector orthogonal to this state is therefore the eigenpolarization of interest for the reflection process. Except in cases having both birefringence and dichroism, a Jones matrix $M_t$ has two orthogonal eigenvectors [38]. While the metasurface in Fig. 1(e) has a rectangular lattice composed of elliptical inclusions, its birefringence is small due to the orthogonal orientations of the ellipses. We therefore expect, irrespective of the choice of $\Delta\alpha$ and $\alpha$, to find an elliptical eigenpolarization with near-unity reflectance, while the orthogonal eigenpolarization has near-unity transmittance.

For instance, Fig. 2(a) maps the peak reflectance $R$ for the metasurface of Fig. 1(e) with $\alpha = 47^\circ$ and $\Delta\alpha = 31^\circ$ as a function of the ellipticity parameters of the polarization state $2\chi$ and $2\psi$, representing the latitude and longitude on the Poincare sphere, respectively. We find a peak with near-unity reflectance (marked with a red dot) and with near-unity transmittance (marked with a blue dot) at the resonant frequency. Figure 2(b) shows the reflectance spectra for these two extremal cases, demonstrating a sharp Fano resonance for one of the eigenpolarizations, and minimal response for the orthogonal polarization state. The inset shows that the eigenpolarization maintains its polarization state, including its handedness, upon reflection. The values of $2\chi$ and $2\psi$ may be determined for every choice of $\Delta\alpha$ and $\alpha$ using a similar procedure, and are shown in Figs. 2(c,d). Hence, by varying the metasurface geometry we can span the entire Poincare sphere, with $2\chi$ controlled primarily by $\Delta\alpha$ and $2\psi$ controlled primarily by $\alpha$.

Next, we study metasurfaces designed for RCP illumination where the elliptical eigenpolarization is spatially varied across the device. We begin by building a library of meta-units (the building blocks of a spatially varying metasurface) capable of controlling both the amplitude and phase of the resonantly reflected RCP light, comparable to the results in Ref. [39] but applied only across the designed bandwidth of the q-BIC. For RCP light incident on a device with eigenpolarization $|e\rangle$, the complex amplitude of the RCP component of the reflected wave is

$$E_R = \langle e|R\rangle^2 = \frac{1}{2}(1+\sin(2\chi))e^{2i\psi} . \quad (1)$$

The remaining states (LCP in reflection, LCP and RCP in transmission) may be determined in a similar manner [40]. It is clear to see that the amplitude of reflected RCP light varies from $0$ to $1$ as $2\chi$ varies from $-\pi/2$ to $\pi/2$ (the poles of the Poincare sphere) while the phase is simply $\Phi = 2\psi$. According to the relation between $2\psi$ and $\alpha$ in Fig. 2(d), the phase of the reflected RCP light varies as $\Phi \sim 4\alpha$, which is twice the conventional geometric phase.

Full-wave simulations confirm this picture: Figures 3(a,b) show, respectively, the reflectance spectra for RCP and LCP incident light for fixed $\alpha = 70^{\circ}$ but varying $\Delta\alpha$, demonstrating complete control of the amplitude of reflected light by varying the latitude of the eigenpolarization. Figures 3(c,d) map the amplitude and phase of reflected RCP light as a function of $\alpha$ and $\Delta\alpha$, showing complete coverage of the two parameters, closely following the coverage of the Poincaré sphere in Figs. 2(c,d). Finally, Figs. 3(e,f) invert the information in the previous panels to realize a look-up table prescribing the required geometries to obtain the desired combinations of amplitude and phase. This tool is ideal to apply the concept of geometric phase to a metasurface with slow variations of amplitude and phase in its transverse plane to synthesize a wavefront transformation. Compared to usual metasurface approaches, the proposed chiral q-BICs enable highly efficient transformations with controllable linewidths.

Figure 4 demonstrates, as an example, a phase-gradient device anomalously reflecting incoming RCP light with near-unity efficiency only at resonance (leaving all non-resonant light specularly transmitted). We choose a sub-library of Figs. 3(e,f) with maximal amplitude and complete phase coverage, seen in Fig. 4(a). Figure 4(b) shows that the resonant wavelength varies less than the linewidth of the resonance across all members of this sub-library, and reports the values of $\Delta\alpha$ that maximize the amplitude for each choice of $\alpha$. This sub-library provides a spatial phase gradient by varying $\alpha$; the resulting geometry is seen in Fig. 4(c), overlaid with the amplitude and phase of the q-BIC at the interface between the PCS and the substrate when excited by RCP light. The far field for each wavelength is shown in Figs. 4(d,e) for RCP and LCP light, demonstrating near-unity anomalous reflection when RCP light is incident, and near-unity specular transmission when LCP light is incident.

As discussed in [19], a phase-gradient metasurface realized with achiral perturbations, as in Fig. 1(a), may be realized for circularly polarized light by spatially varying $\alpha$. If the device has symmetry about the horizontal plane at $z = 0$, i.e., $\Delta\alpha = 0$, the eigenpolarization necessarily has $\chi = 0$. Equation (1) shows that the peak efficiency for anomalously reflected light is capped at $25\%$ in this case; the analogous equations for the remaining three components show that the device equally splits light into four ports at the resonant frequency: specular transmission (RCP) and reflection (LCP), and anomalous transmission (LCP) and reflection (RCP) [40]. In this achiral scenario, linear momentum conservation precludes all but two diffraction orders for each spin as interference pathways. Here, in contrast, a chiral device with circularly polarized eigenpolarizations ( $\chi = \pm\pi / 4$ ) eliminates the two interference pathways associated with the non-resonant spin. What remains is an effective two-port system supporting a sharp Fano resonance with near-unity reflectance to an arbitrary angle (the $m = \pm 2$ diffraction order).

More complex wavefront shaping, such as focusing or orbital angular momentum generation, is possible following similar principles. The library in Fig. 3 generally enables phase-amplitude holography encoded in a chiral q-BIC that can be optically reconstructed only using the correct input polarization and wavelength; all non-resonant wavelengths will see a slab acting as a weakly perturbed effective medium. In total, a chiral q-BIC enables engineering with large flexibility the Fano resonance Q-factor, resonant wavelength, and phase-amplitude response simultaneously by tuning the magnitude of the perturbation, unperturbed lattice geometry, and orientation angles of the chiral structure, respectively. In the spatial domain, the phase and amplitude control translates into control of the local diffraction angles and diffraction efficiencies, respectively. The approach outlined here therefore represents the ultimate control of a generalized Fano resonance with spatially tailored dark modes. We consider this to form an exciting platform for applications such as augmented reality and secure optical communications, which enable a broad transparent response unless interrogated in a very narrow spectral range with a specific eigenpolarization of choice [40].

In conclusion, here we have extended the concept of q-BICs to chiral symmetry perturbations, expanding the range of available eigenpolarization states of a q-BIC to the entire Poincare sphere. When the eigenpolarization is circularly polarized, the device supports a Fano resonance exhibiting full circular dichroism, with the eigenpolarization reflected (and the orthogonal polarization transmitted) with near-unity efficiency and arbitrarily sharp frequency selectivity. When the incident light is fixed to be circularly polarized, varying the eigenpolarization of the device in space tunes the amplitude and phase of reflected light of the same spin. Hence, by applying geometric phase considerations our findings enable metasurfaces encoding spatial phase profiles with near-unity amplitude, promising for highly spectrally and spin selective optical wavefront shaping with a bandwidth tunable by the degree of asymmetric perturbation.

This work was supported by DARPA, ONR, and AFOSR.

*To whom correspondence should be addressed: aalu@gc.cuny.edu, ny2214@columbia.edu.

## Figures

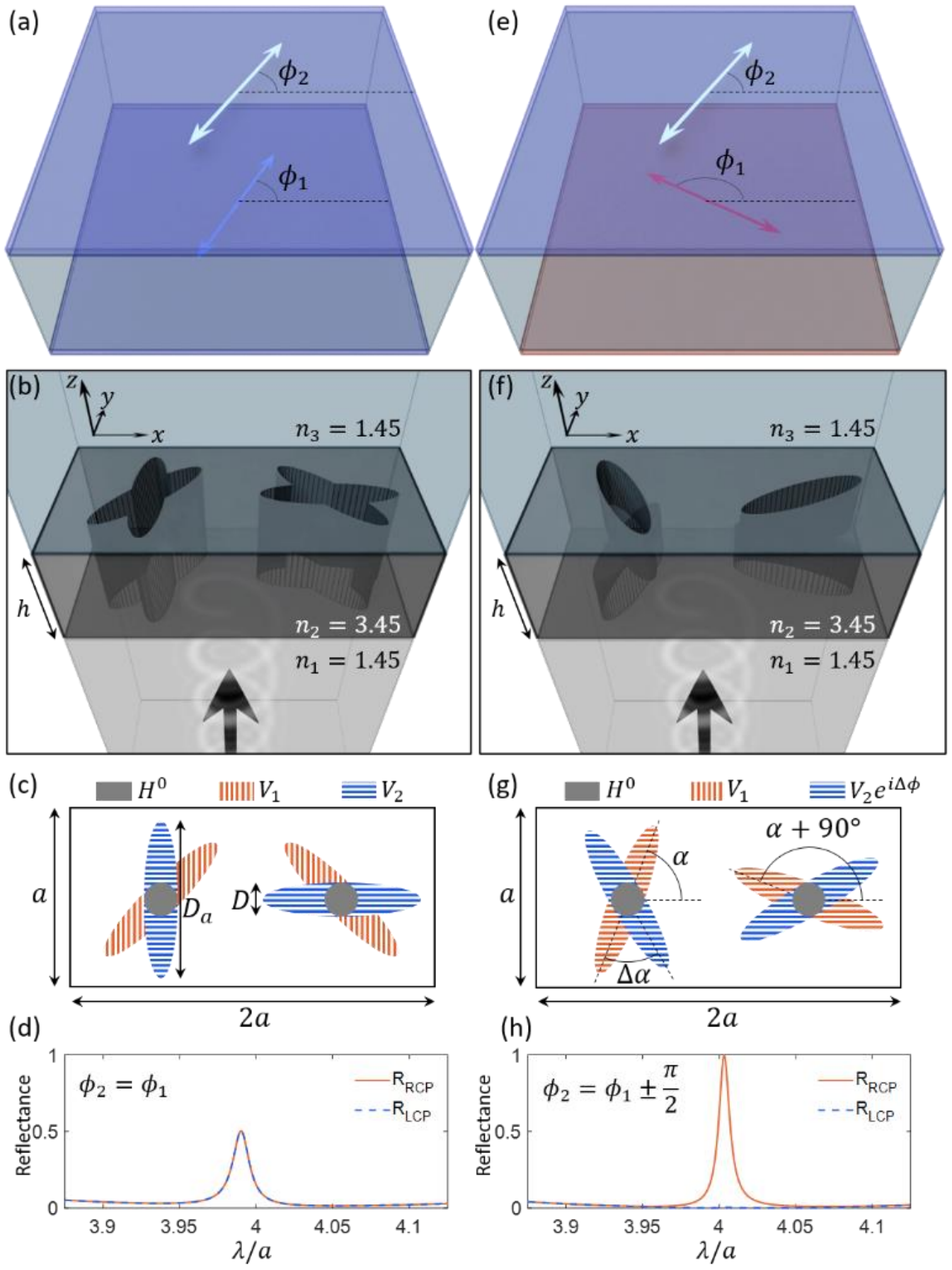

**Figure 1.** Chiral q-BICs. (a,e) Homogenized models of metasurfaces supporting q-BICs: two interfaces scatter optical energy to linearly polarized free-space light with polarization angles $\phi_1$ and $\phi_2$, respectively; optical chirality is achievable only when $\phi_1 \neq \phi_2$. (b,f) Unit cell of sample implementations of achiral and chiral metasurfaces composed of two perturbations, $V_1$ and $V_2$, shown in (c,g). In the chiral version, $V_2$ has a relative phase factor $e^{i\Delta\phi}$ according to its vertical displacement from $V_1$. (d,h) Reflectance spectra for RCP and LCP light near a quasi-BIC when $h = 1.25a$, $D = 0.2a$, and $D_a = 0.85a$, showing no circular dichroism in the achiral case and full circular dichroism in the chiral case when $\alpha = 70^{\circ}$ and $\Delta\alpha = 50^{\circ}$.

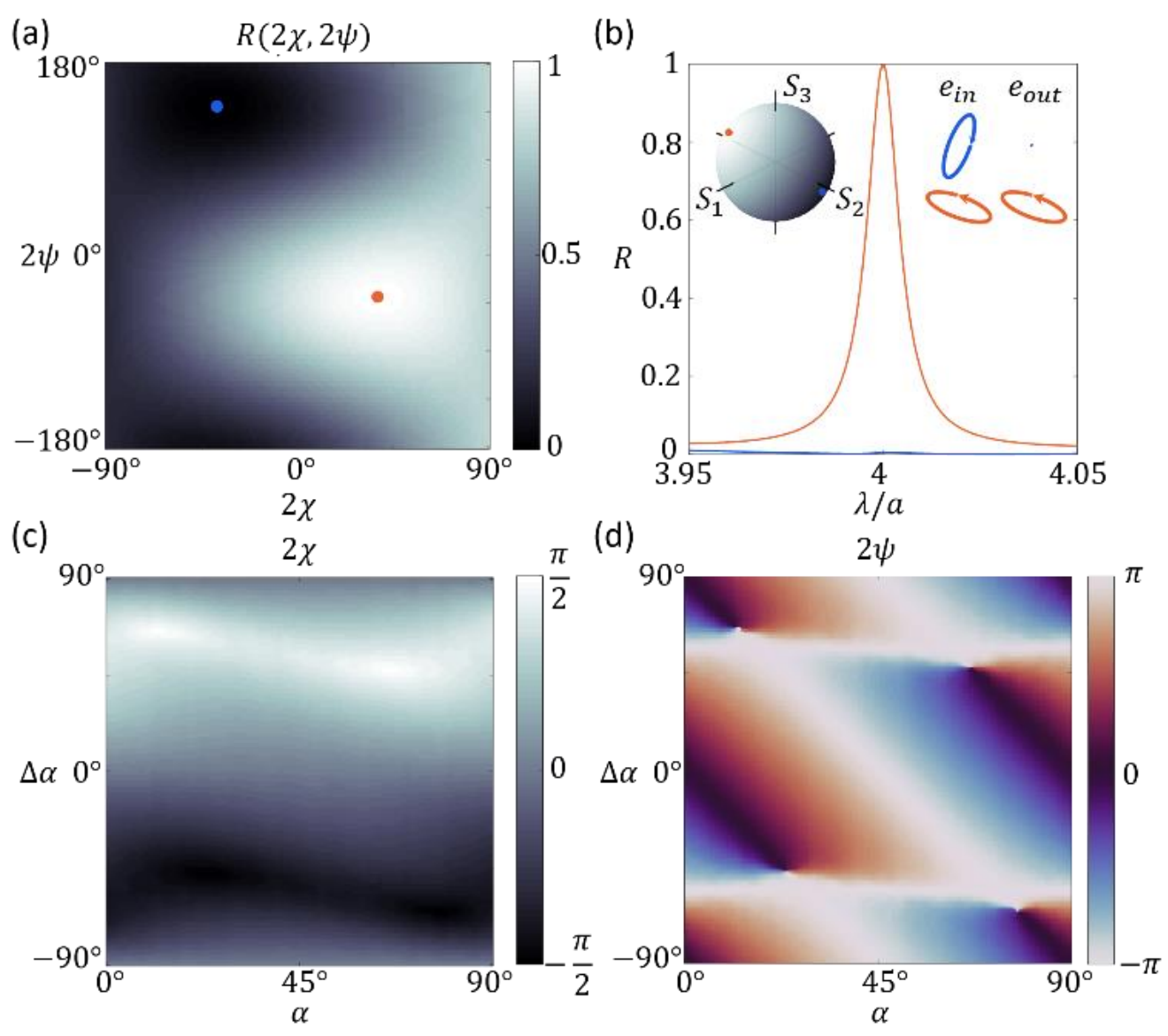

**Figure 2.** Fano resonances with arbitrary elliptical eigenpolarizations. (a) Map of reflectance for a device with $\alpha = 47^{\circ}$ and $\Delta\alpha = 31^{\circ}$ as a function of input polarization state (characterized by the latitude, $2\chi$, and longitude, $2\psi$, on the Poincare sphere). (b) Reflectance spectra for the extremal cases in (a) (marked by red and blue dots), with the insets showing the their positions on the Poincare sphere and the eigenpolarization states, *e*, that are maximally and minimally resonant. (c,d) Map of the coverage of the Poincare sphere for the maximally resonant elliptical eigenpolarization as a function of $\alpha$ and $\Delta\alpha$.

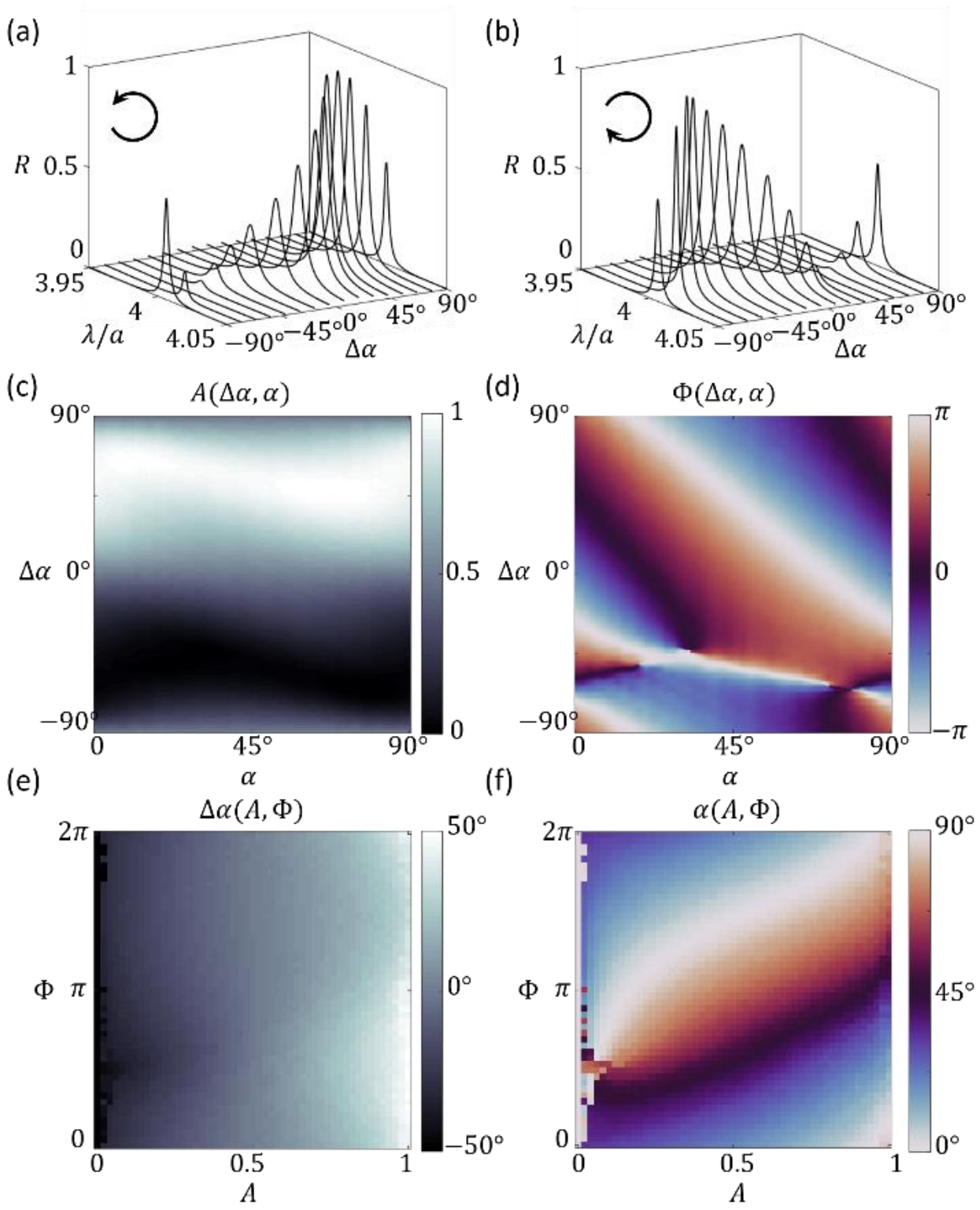

**Figure 3.** Full control over amplitude and phase of resonantly reflected light. (a,b) Reflectance, *R*, for RCP and LCP light with $\alpha = 70^{\circ}$ and varying $\Delta\alpha$, showing that the peak reflectance may vary from 0 to 1 depending on the chirality of the structure. (c,d) Amplitude, *A*, and phase, $\Phi$ at the resonant wavelength for reflected RCP light as a function of $\alpha$ and $\Delta\alpha$. (e,f) Look-up design table specifying the values of $\alpha$ and $\Delta\alpha$ required to get a desired combination of *A* and $\Phi$.

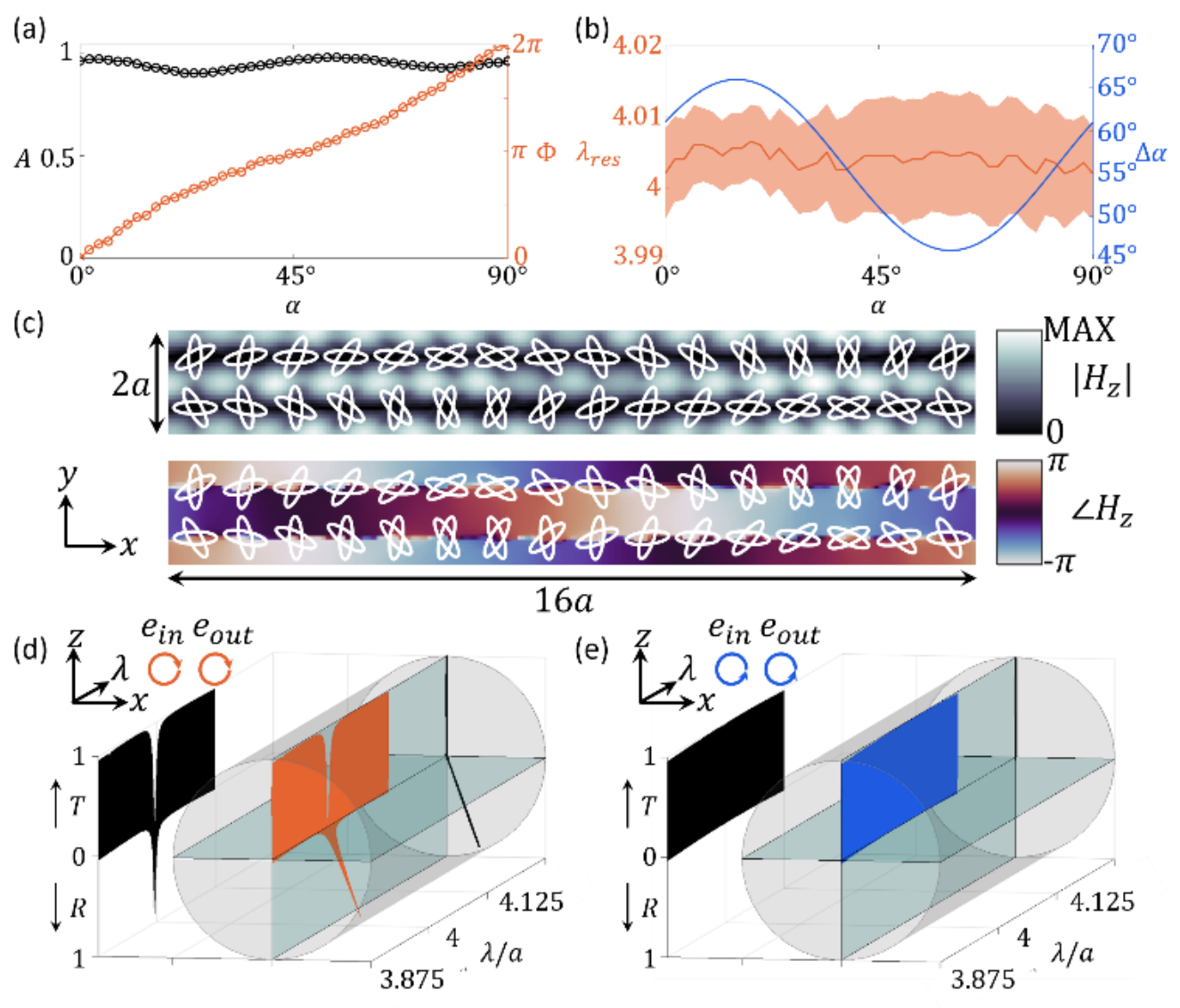

**Figure 4.** Chiral phase-gradient resonant metasurface with near-unity diffraction efficiency. (a) Sub-library of Fig. 3(c,d) with near-unity amplitude and phase varying over $2\pi$ . (b) Resonant wavelength, $\lambda_{res}$ , for each element in (a), with the shaded region depicting the full width at half max of the resonance. Also depicted is the value of $\Delta\alpha$ used for each $\alpha$ in (a). (c) Mode profile on resonance for a phase-gradient device implemented using the library in (a). The ellipses denote the geometry of the structure. (d) Far field projection for RCP incidence at each wavelength for the device shown in (c), demonstrating near-unity diffraction efficiency ( $96.1\%$ ) to the anomalously reflected angle ( $20.17^{\circ}$ in the substrate material). The spin is preserved in transmission and reflection. (e) Far-field projection for LCP incidence at each wavelength for the device in (c), showing no resonance for this eigenpolarization.

**Supplementary Materials for**

# Chiral Quasi-Bound States in the Continuum

Adam Overvig[1,2], Nanfang Yu[2*], and Andrea Alù[1*]

[1]Photonics Initiative, Advanced Science Research Center at the Graduate Center of the City University of New York

[2]Department of Applied Physics and Applied Mathematics, Columbia University

## 1. Selection Rules

To determine the impact of the perturbation on the high symmetry photonic crystal slab, we must apply the relevant selection rules to the modes supported in the perturbed metasurface. The selection rules (retrieved from Ref. [13] in the main text) for each $V_1$ and $V_2$ (reproduced here in Fig. S1(a,b)) are rules that forbid or allow excitation of modes according to the modes' in-plane symmetries. The perturbed device supports modes with momentum matched to normally incident light (the $\Gamma$-point) of two kinds: those that existed at the $\Gamma$-point in the unperturbed slab, and those that are folded from the $M$ -point by the dimerization. The BICs supported by the unperturbed photonic crystal slab at the $\Gamma$ -point remain bound in the continuum for all combinations of perturbations $V_1$ and $V_2$ studied here, and so are omitted.

For the modes folded from the $M$ -point of the unperturbed lattice, there are four distinct combinations of even and odd reflection symmetries. These are tracked in the language of Group Theory by the irreducible representations $A_1, A_2, B_1, B_2$ of the point group $C_{2v}$ of the $M$ -point ( Fig. S1(e) for example mode profiles). While modes that transform like each of these irreducible representations are supported by the device, the selection rules (Fig. S1(c,d)) show that only the $A_1$ and $A_2$ modes couple to free-space for $V_1$ and $V_2$. In other words, coupling to $B_1$ and $B_2$ modes constitutes a class of symmetry-protected BICs, while coupling to $A_1$ and $A_2$ modes corresponds to q-BICs. The $A_2$ and $B_2$ modes are typically higher-order modes, spectrally separated from the lowest frequency modes supported by the structure, which are typically $A_1$ and $B_1$ modes. Since coupling to $B_1$ modes is forbidden, we are left with only $A_1$ modes, which couple to $y$ in the presence of $V_1$ and to $x$ in the presence of $V_2$. We therefore may expect simple spectra (i.e., isolated $A_1$ modes) near the frequency of the fundamental mode, and we emphasize that this symmetry-based argument is valid for a wide range of suitably scaled structures describable by perturbations matching the symmetries of $V_1$ and $V_2$.

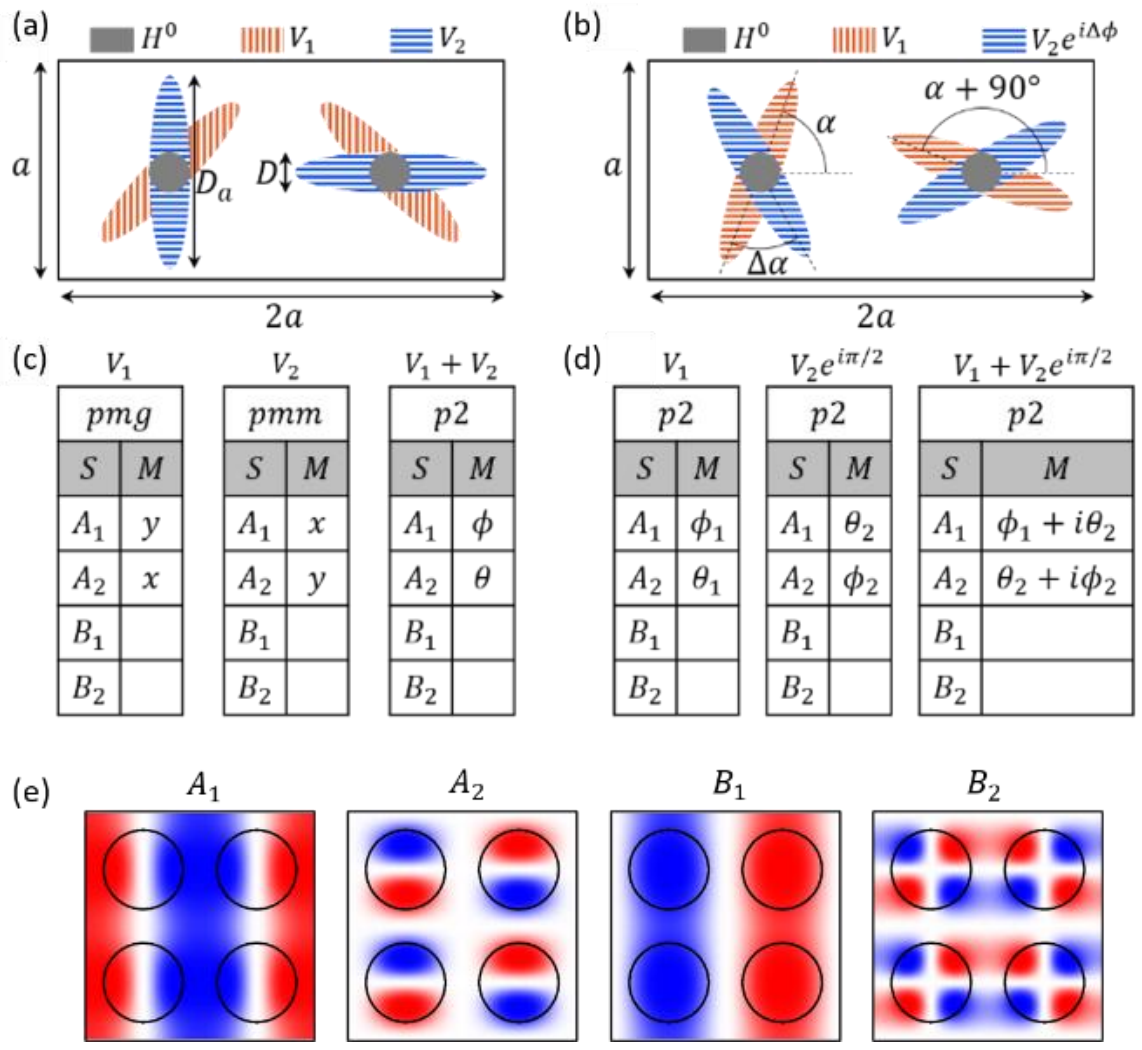

Fig. S1. Selection rules for the achiral and chiral devices in the main text. (a,b) Schematics of the perturbed structures, reproduced from the main text. (c,d) Selection rules enumerated for the zone folded modes, determined by applying Group Theory principles given the space group (e.g., *pmg*) and irreducible representations, $S$. Entries specify the eigenpolarizations of the q-BICs in each case, while blank entries mean that the modes remain bound in the contiinum. (e) Example modal profiles for each irreducible representation. The $A_1$ mode depicted here is the quasi-BIC explored in the main text.

## 2. Temporal Coupled Mode Theory

A useful phenomenological model for q-BICs is a Temporal Coupled Mode Theory (TCMT). We begin by writing down the generic form of the scattering matrix for the device shown in Fig. S2, which is a slab suspended in air that scatters light at the interfaces due to dimers with separately controlled selection rules. The index of refraction, $n$, and height of the slab, $H$, will determine the Fresnel coefficients, $r$ and $t$, where for normal incidence

$$t = \sqrt{1-r^2}\,. \tag{1.1}$$

Then, for fields of the form

$$E = \begin{bmatrix} E_{x,1} \\ E_{x,2} \\ E_{y,1} \\ E_{y,2} \end{bmatrix} \tag{1.2}$$

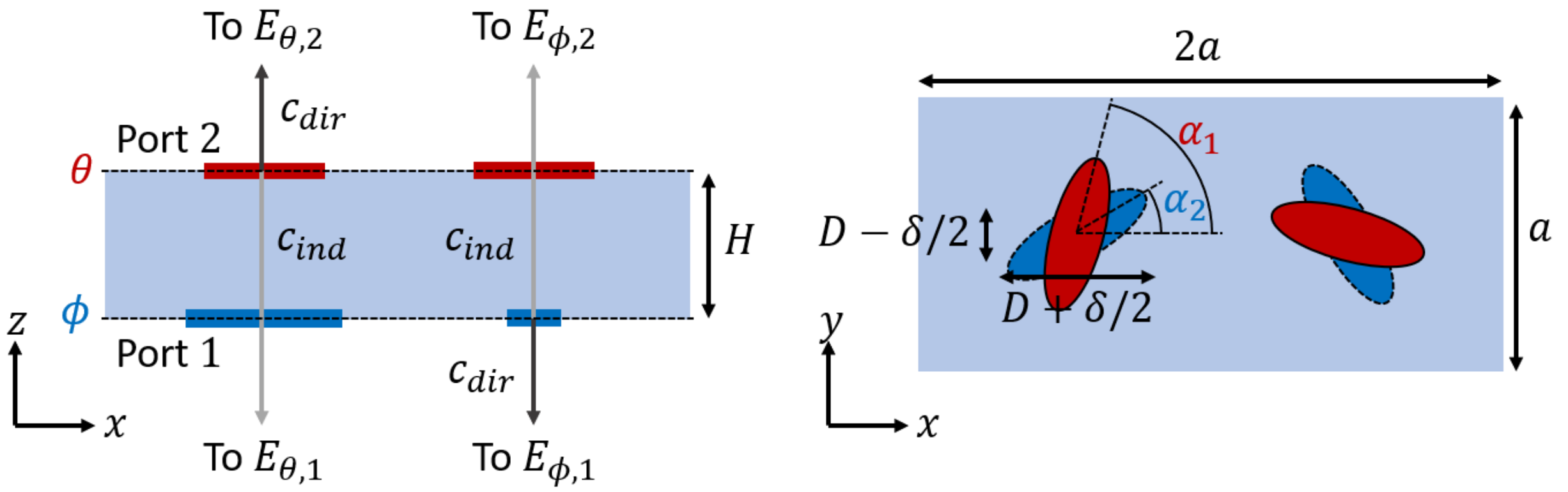

Fig. S2. Schematics of a unit cell of the device for the development of the temporal coupled mode theory. (Left) Side view, showing two interfaces scattering to linear polarization to two ports. (Right) Top view, showing the in-plane orientation angles of the ellipses.

where $E_{m,p}$ describe the fields polarized in the $m$ direction corresponding to port $p$, the background scattering matrix is:

$$C = \begin{bmatrix} r & it & 0 & 0 \\ it & r & 0 & 0 \\ 0 & 0 & r & it \\ 0 & 0 & it & r \end{bmatrix}. \tag{1.3}$$

The overall scattering matrix including the q-BIC is

$$S = C + \frac{|d\rangle\langle d^*|}{i(\omega-\omega_0)+1/\tau} \tag{1.4}$$

where $\omega$ is the angular frequency, $\omega_0$ is the resonant angular frequency, $\tau = Q/\omega_0$ is the optical lifetime of the q-BIC with a Q-factor $Q$, and

$$|d\rangle = \begin{bmatrix} d_1 \\ d_2 \\ d_3 \\ d_4 \end{bmatrix} \tag{1.5}$$

are the coupling coefficients of the q-BIC to the four external fields of Eqn. (1.2). The overall scattering matrix can be written explicitly as:

$$S = \begin{bmatrix} r & it & 0 & 0 \\ it & r & 0 & 0 \\ 0 & 0 & r & it \\ 0 & 0 & it & r \end{bmatrix} + \frac{1}{i(\omega-\omega_0)+1/\tau} \begin{bmatrix} d_1d_1 & d_1d_2 & d_1d_3 & d_1d_4 \\ d_2d_1 & d_2d_2 & d_2d_3 & d_2d_4 \\ d_3d_1 & d_3d_2 & d_3d_3 & d_3d_4 \\ d_4d_1 & d_4d_2 & d_4d_3 & d_4d_4 \end{bmatrix}. \tag{1.6}$$

Here, all of the parameters may be controlled explicitly within the q-BIC framework: the Fresnel coefficients are controlled by $n$ and $H$, $\omega_0$ is controlled by the period and geometry of the unperturbed PCS, and $\tau = Q/\omega_0 \propto 1/\delta^2$ where $\delta$ characterizes the magnitude of the symmetry-breaking perturbation. What remains is to parameterize the elements of $|d\rangle$ in terms of geometric parameters, and to apply the relevant physical constraints.

In the general case, we know from the selection rules that the two interfaces may each couple the q-BIC to free-space to any linear polarization state. We call $\phi$ the polarization angle coupled by the perturbation at the $p=1$ interface and $\theta$ the polarization angle coupled by the perturbation at the $p=2$ interface. Then, the contribution to the elements of $|d\rangle$ may be decomposed as:

$$\begin{aligned} d_1 &= c_{1\phi}\cos(\phi) + c_{1\theta}\cos(\theta) \\ d_2 &= c_{2\phi}\cos(\phi) + c_{2\theta}\cos(\theta) \\ d_3 &= c_{1\phi}\sin(\phi) + c_{1\theta}\cos(\theta) \\ d_4 &= c_{2\phi}\sin(\phi) + c_{2\theta}\cos(\theta) \end{aligned} \tag{1.7}$$

where the coefficients $c_{p\gamma}$ represent the coupling to port $p$ from the scatterer producing polarization $\gamma$. These coefficients in the general case may be independent. For instance, we expect $c_{1\theta}$ to differ from $c_{1\phi}$ because the former represents direct scattering to port 1 while the latter represents indirect scattering from the opposite interface through the slab. Likewise, $c_{2\theta}$ is direct and $c_{2\phi}$ is indirect. The symmetry of the unperturbed structure demands that the direct coefficients are equivalent to each other for the same perturbation strength, and likewise for the two indirect coefficients. This leaves two coefficients $c_{dir}$ and $c_{ind}$:

$$\begin{aligned} c_{2\theta} &= c_{1\phi} \equiv c_{dir} \\ c_{1\theta} &= c_{2\phi} \equiv c_{ind} \end{aligned}. \tag{1.8}$$

We may expect in general that both the amplitude and phase of the direct scattering coefficient, $c_{dir}$, differ from the indirect scattering coefficient, $c_{ind}$. That is,

$$c_{ind} = ge^{i\Delta\Phi}c_{dir} \tag{1.9}$$

where $g$ is the amplitude ratio and $\Delta\Phi$ is the phase difference. Since the direct and indirect pathways differ simply by propagation through the slab, it is natural to set $\Delta\Phi = nH\frac{\omega}{c_0}$, where $c_0$ is the speed of light in vacuum. This applies for q-BICs that decay symmetrically, but for q-BICs that decay anti-symmetrically, Eqn. (1.9) should be adjusted to

$$c_{ind} = -ge^{i\Delta\Phi}c_{dir}. \tag{1.10}$$

For simplicity, we will assume here that we are working with a symmetrically decay mode, but the results are easily generalized to anti-symmetrically decaying modes (such as in the main text).

Next, we apply the conditions required by energy conservation,

$$\langle d | d \rangle = \frac{2}{\tau} \tag{1.11}$$

and time-reversal symmetry,

$$C\left|d^*\right\rangle = -\left|d\right\rangle, \tag{1.12}$$

to determine $c_{dir}$ and $g$. From Eqn. (1.11) we have

$$\left|d_1\right|^2+\left|d_2\right|^2+\left|d_3\right|^2+\left|d_4\right|^2=\frac{2}{\tau} \tag{1.13}$$

which, using Eqn. (1.7) and Eqn. (1.9), leads to

$$\left|c_{dir}\right|^2=\frac{1/\tau}{1+g^2+2g\cos(\phi-\theta)\cos(\Delta\Phi)}\,. \tag{1.14}$$

Therefore, we have

$$c_{dir}=\sqrt{\frac{1/\tau}{1+g^2+2g\cos(\phi-\theta)\cos(\Delta\Phi)}}e^{i\Phi_{dir}}\,, \tag{1.15}$$

and what remains is to find the phase $\Phi_{dir}$ and amplitude ratio $g$ . These are easily found by applying Eqn. (1.12), which enforces

$$rd_1^*+itd_2^*=-d_1\,. \tag{1.16}$$

Using Eqn. (1.7) and Eqn. (1.15), we have

$$h\equiv\frac{(r+itge^{-i\Delta\Phi})\cos(\phi)+(rge^{-i\Delta\Phi}+it)\cos(\theta)}{\cos(\phi)+\cos(\theta)ge^{i\Delta\Phi}}=-e^{2i\Phi_{dir}}\,. \tag{1.17}$$

The right-hand side requires that $\left|h\right|=1$, which requires

$$g=\frac{\left|\sqrt{1-r^2\cos(\Delta\Phi)^2}-r\sin(\Delta\Phi)\right|}{\tau}\,. \tag{1.18}$$

Then, the phase $\Phi_{dir}$ is simply given by the known parameters in Eqn. (1.17), and we arrive at

$$c_{dir} = \sqrt{\frac{-h/\tau}{1+g^2+2g\cos(\phi-\theta)\cos(\Delta\Phi)}}\,. \tag{1.19}$$

With Eqn. (1.19) in hand, the elements of $|d\rangle$ are known, and the overall scattering matrix is uniquely determined by the phenomenological parameters $r, \omega_0, \tau, \phi, \theta, \Delta\Phi$, which in turn are controlled by geometric (and material) parameters as

$$\begin{aligned}
&r = \sin(\Delta\Phi)\frac{1}{1+\frac{2in}{(1+n^2)\tan(\Delta\Phi)}}\\
&\omega_0 = f(\epsilon)\\
&\tau \propto 1/\delta^2\\
&\phi \approx 2\alpha_1\\
&\theta \approx 2\alpha_2\\
&\Delta\Phi = nH\omega/c_0
\end{aligned} \tag{1.20}$$

where $f(\epsilon)$ represents some function (with no generalized closed form) of the period, permittivity, and other geometrical parameters of the unperturbed PCS. If an anti-symmetrically decay mode is being studied, Eqn. (1.20) has the substitutions:

$$\begin{aligned}
&r = \sin(\Delta\Phi-\pi)\frac{1}{1+\frac{2in}{(1+n^2)\tan(\Delta\Phi-\pi)}}\\
&\Delta\Phi = nH\omega/c_0+\pi
\end{aligned} \tag{1.21}$$

Readily apparent from Eqn. (1.20) is that the case when $\Delta\Phi = \pi/2$ means that the background scattering has no reflectance, $r = 0$, and linear polarizations $\phi$ and $\theta$ have a phase delay corresponding to a circular polarization. That is to say, when operating at a Fabry-Perot resonance such that the transmission is maximal, the results of the TCMT straightforwardly predict that the circular dichroism is maximal. To see this more explicitly, we may analyze the Jones matrices of the system, as determined by the scattering matrix.

### 3. Eigenvectors of the Jones matrix

With the TCMT in hand, we may now write down explicitly the reflection side Jones matrix as

$$M_r = \begin{bmatrix} S_{11} & S_{13} \\ S_{31} & S_{33} \end{bmatrix} \tag{1.22}$$

where

$$\begin{aligned} S_{11} &= r - f(\cos(\phi) + \cos(\theta) g e^{i\Delta\Phi})^2 \\ S_{13} = S_{31} &= -f(\cos(\phi) + \cos(\theta) g e^{i\Delta\Phi}) \sin(\phi) + \sin(\theta) g e^{i\Delta\Phi} \\ S_{33} &= r - f(\sin(\phi) + \sin(\theta) g e^{i\Delta\Phi})^2 \end{aligned} \tag{1.23}$$

and

$$f = \frac{1}{i(\omega - \omega_0) + 1/\tau} \frac{h/\tau}{1 + g^2 + 2g\cos(\phi - \theta)\cos(\Delta\Phi)}\,. \tag{1.24}$$

We are principally interested in when $\Delta\Phi = \pi/2$, in which case we have

$$M_r = -\frac{1}{2}\begin{bmatrix} (\cos(\phi) + i\cos(\theta))^2 & (\cos(\phi) + i\cos(\theta))(\sin(\phi) + i\sin(\theta)) \\ (\cos(\phi) + i\cos(\theta))(\sin(\phi) + i\sin(\theta)) & (\sin(\phi) + i\sin(\theta))^2 \end{bmatrix}. \tag{1.25}$$

It is readily apparent is that $M_r$ is a symmetric singular matrix, from which we know that it has two orthogonal eigenvectors (eigenpolarizations), at least one of which has an eigenvalue of 0.

That is, one of the eigenpolarizations reflects identically to the background, $r = 0$, and is therefore non-resonant, leaving the other eigenpolarization to be resonant. In particular, the eigenpolarizations are

$$
\begin{aligned}
|e_1\rangle &= \begin{bmatrix} \cos(\phi) + i\cos(\theta) \\ \sin(\phi) + i\sin(\theta) \end{bmatrix} \\
|e_2\rangle &= \begin{bmatrix} -(\sin(\phi) + i\sin(\theta)) \\ \cos(\phi) + i\cos(\theta) \end{bmatrix}
\end{aligned}
\tag{1.26}
$$

where $|e_1\rangle$ is the resonant eigenpolarization with (non-normalized) eigenvalue $v_1 = i\cos(\phi - \theta)$ and $|e_2\rangle$ is the orthogonal, non-resonant eigenpolarization with eigenvalue $v_2 = 0$.

With a change of variables, we may put the resonant eigenpolarization in a more suggestive form. With

$$
\begin{aligned}
\psi &= \frac{\phi + \theta}{2} \\
\chi &= \frac{\phi - \theta}{2}
\end{aligned}
\tag{1.27}
$$

we may write

$$
|e_1\rangle = \begin{bmatrix} \cos(\psi) & -\sin(\psi) \\ \sin(\psi) & \cos(\psi) \end{bmatrix} \begin{bmatrix} \cos(\chi) \\ i\sin(\chi) \end{bmatrix}, \tag{1.28}
$$

which is a well-known parameterization of an arbitrary polarization state whose latitude and longitude on the Poincare sphere are $2\psi$ and $2\chi$ respectively. Since the selection rules give

$$
\begin{aligned}
\phi &\approx 2\alpha_1 \\
\theta &\approx 2\alpha_2
\end{aligned}
\tag{1.29}
$$

we may achieve any elliptical eigenpolarization according to

$$\begin{aligned} \psi &\approx \alpha_1 + \alpha_2 \\ \chi &\approx \alpha_2 - \alpha_1 \end{aligned} \tag{1.30}$$

When $2\chi = \pm\pi/2$ (that is, $\alpha_2 = \alpha_1 \pm \pi/4$), we have circularly polarized light, and the resonant polarization is

$$|e_1\rangle = \frac{1}{\sqrt{2}} \begin{bmatrix} \cos(2\alpha_1 \pm \pi/4) & -\sin(2\alpha_1 \pm \pi/4) \\ \sin(2\alpha_1 \pm \pi/4) & \cos(2\alpha_1 \pm \pi/4) \end{bmatrix} \begin{bmatrix} 1 \\ \pm i \end{bmatrix} = \frac{e^{-i\pi/4} e^{\mp 2i\alpha_1}}{\sqrt{2}} \begin{bmatrix} 1 \\ \pm i \end{bmatrix}. \tag{1.31}$$

Which is to say, the resonant eigenpolarization is circularly polarized with phase that varies with $\alpha_1$. We note that this is not the reflection phase, but a portion of it: the full picture of the reflection requires projecting the incident polarization state onto the eigenpolarization, and then decomposing the reflected light into an orthogonal basis. This is detailed in the following section.

### 4. Derivation of Equation (1) in the main text

We now explicitly show the derivation of Eq. (2) of the main text. Consider an RCP state traveling in the positive $z$ direction, $|R\rangle$. The device will reflect the portion of the light that is the eigenpolarization, $|e\rangle$, to the state $|e^*\rangle$, where the complex conjugate accounts for the reversal of direction. In other words, the reflected state is given by the projection operator $P_e = |e^*\rangle\langle e|$. Next, the circular polarization filter leaves only the portion of the reflected light with that is RCP travelling in the negative $z$ direction, $\left|R^*\right\rangle$ , which is given by the projection operator $P_{R^*}=\left|R^*\right\rangle\left\langle R^*\right|$. The final state of interest exiting the device is therefore

$$P_{R^*}P_e\left|R\right\rangle=\left|R^*\right\rangle\left\langle R^*\middle|e^*\right\rangle\left\langle e\middle|R\right\rangle=\left\langle e\middle|R\right\rangle^2\left|R^*\right\rangle, \tag{1.32}$$

which has a complex amplitude that varies with the eigenpolarization. Defining the Jones vector of RCP light

$$\left|R\right\rangle=\frac{1}{\sqrt{2}}\begin{bmatrix}1\\ i\end{bmatrix}, \tag{1.33}$$

and the Jones vector of an elliptical state characterized by elliptical parameters $\chi$ and $\psi$ :

$$\left|e\right\rangle=\begin{bmatrix}\cos(\psi) & -\sin(\psi)\\ \sin(\psi) & \cos(\psi)\end{bmatrix}\begin{bmatrix}\cos(\chi)\\ i\sin(\chi)\end{bmatrix}=\begin{bmatrix}\cos(\psi)\cos(\chi)-i\sin(\psi)\sin(\chi)\\ \sin(\psi)\cos(\chi)+i\cos(\psi)\sin(\chi)\end{bmatrix} \tag{1.34}$$

we then evaluate the quantity $\left\langle e\middle|R\right\rangle$:

$$\begin{aligned}\left\langle e\middle|R\right\rangle&=\frac{1}{\sqrt{2}}\begin{bmatrix}\cos(\psi)\cos(\chi)+i\sin(\psi)\sin(\chi) & \sin(\psi)\cos(\chi)-i\cos(\psi)\sin(\chi)\end{bmatrix}\begin{bmatrix}1\\ i\end{bmatrix}\\ &=\frac{1}{\sqrt{2}}\Big(\big(\cos(\psi)\cos(\chi)+\cos(\psi)\sin(\chi)\big)+i\big(\sin(\psi)\sin(\chi)+\sin(\psi)\cos(\chi)\big)\Big)\\ &=\frac{1}{\sqrt{2}}\big(\cos(\chi)+\sin(\chi)\big)e^{i\psi}\end{aligned} \tag{1.35}$$

Therefore, the coefficient $\left\langle e\middle|R\right\rangle^2$ is:

$$
\begin{aligned}
\left\langle e \middle| R \right\rangle^2 &= \frac{1}{2}\left(\cos(\chi)+\sin(\chi)\right)^2 e^{2i\psi} = \frac{1}{2}\left(\cos^2(\chi)+\sin^2(\chi)+2\cos(\chi)\sin(\chi)\right)e^{2i\psi} \\
&= \frac{1}{2}\left(1+\sin(2\chi)\right)e^{2i\psi}
\end{aligned}
\tag{1.36}
$$

Similarly, we determine remaining three components (LCP in reflection and LCP and RCP in transmission). Starting with the LCP in reflection, we have

$$
P_{L^*}P_e\left|R\right\rangle = \left|L^*\right\rangle\left\langle L^* \middle| e^*\right\rangle\left\langle e \middle| R\right\rangle = \left\langle e \middle| L\right\rangle\left\langle e \middle| R\right\rangle\left|L^*\right\rangle . \tag{1.37}
$$

Proceeding as before, we have

$$
\begin{aligned}
\left\langle e \middle| L \right\rangle &= \frac{1}{\sqrt{2}}\left[\cos(\psi)\cos(\chi)+i\sin(\psi)\sin(\chi) \quad \sin(\psi)\cos(\chi)-i\cos(\psi)\sin(\chi)\right]\begin{bmatrix} 1 \\ -i \end{bmatrix} \\
&= \frac{1}{\sqrt{2}}\left(\left(\cos(\psi)\cos(\chi)-\cos(\psi)\sin(\chi)\right)+i\left(\sin(\psi)\sin(\chi)-\sin(\psi)\cos(\chi)\right)\right) \\
&= \frac{1}{\sqrt{2}}\left(\cos(\chi)-\sin(\chi)\right)e^{-i\psi},
\end{aligned}
\tag{1.38}
$$

resulting in a final state with a phase term:

$$
\begin{aligned}
\left\langle e \middle| L\right\rangle\left\langle e \middle| R\right\rangle &= \frac{1}{2}\left(\cos(\chi)-\sin(\chi)\right)e^{-i\psi}\left(\cos(\chi)+\sin(\chi)\right)e^{i\psi} \\
&= \frac{1}{2}\left(\cos^2(\chi)-\sin^2(\chi)\right) = \frac{1}{2}\cos(2\chi)
\end{aligned}
\tag{1.39}
$$

Next, we determine the transmitted RCP component,

$$
P_R P_{e_t}\left|R\right\rangle = \left|R\right\rangle\left\langle R \middle| e_t\right\rangle\left\langle e_t \middle| R\right\rangle = \left|\left\langle e_t \middle| R\right\rangle\right|^2\left|R\right\rangle, \tag{1.40}
$$

where we construct the polarization orthogonal to the reflected eigenpolarization:

$$\left|e_t\right\rangle = \begin{bmatrix} \sin(\psi)\cos(\chi) - i\cos(\psi)\sin(\chi) \\ -\cos(\psi)\cos(\chi) - i\sin(\psi)\sin(\chi) \end{bmatrix}. \tag{1.41}$$

Using this, we may apply the simplification

$$\begin{aligned} \left\langle e_t \middle| R \right\rangle &= \frac{1}{\sqrt{2}}\left[\sin(\psi)\cos(\chi) + i\cos(\psi)\sin(\chi) \quad -\cos(\psi)\cos(\chi) + i\sin(\psi)\sin(\chi)\right]\begin{bmatrix} 1 \\ i \end{bmatrix} \\ &= \frac{1}{\sqrt{2}}\Big(\big(\sin(\psi)\cos(\chi) - \sin(\psi)\sin(\chi)\big) + i\big(\cos(\psi)\sin(\chi) - \cos(\psi)\cos(\chi)\big)\Big) \\ &= \frac{1}{\sqrt{2}}\big(\cos(\chi) - \sin(\chi)\big)e^{-i\psi}, \end{aligned} \tag{1.42}$$

meaning that the amplitude is

$$\left|\left\langle e_t \middle| R \right\rangle\right|^2 = \frac{1}{2}\big(\cos(\chi) - \sin(\chi)\big)^2 = \frac{1}{2}\big(1 - \sin(2\chi)\big). \tag{1.43}$$

Finally, we determine the transmitted LCP component,

$$P_R P_{e_t}\left|R\right\rangle = \left|L\right\rangle\left\langle L \middle| e_t\right\rangle\left\langle e_t \middle| R\right\rangle = \left\langle e_t \middle| L\right\rangle^* \left\langle e_t \middle| R\right\rangle \left|L\right\rangle, \tag{1.44}$$

Using this, we achieve a similar simplification,

$$\begin{aligned} \left\langle e_t \middle| L \right\rangle &= \frac{1}{\sqrt{2}}\left[\sin(\psi)\cos(\chi) + i\cos(\psi)\sin(\chi) \quad -\cos(\psi)\cos(\chi) + i\sin(\psi)\sin(\chi)\right]\begin{bmatrix} 1 \\ -i \end{bmatrix} \\ &= \frac{1}{\sqrt{2}}\Big(\big(\sin(\psi)\cos(\chi) + \sin(\psi)\sin(\chi)\big) + i\big(\cos(\psi)\sin(\chi) + \cos(\psi)\cos(\chi)\big)\Big) \\ &= \frac{1}{\sqrt{2}}\big(\cos(\chi) + \sin(\chi)\big)e^{i\psi}, \end{aligned} \tag{1.45}$$

meaning that the complex amplitude is

$$\begin{aligned}\langle e_t|L\rangle^* \langle e_t|R\rangle &= \frac{1}{2}\left(\cos(\chi)+\sin(\chi)\right)e^{-i\psi}\left(\cos(\chi)-\sin(\chi)\right)e^{-i\psi} \\ &= \left(\cos^2(\chi)-\sin^2(\chi)\right)e^{-2i\psi} = \cos(2\chi)e^{-2i\psi}.\end{aligned} \tag{1.46}$$

Table S1 summarizes the circularly polarized light both reflected and transmitted, with amplitudes controlled by $2\chi \approx 2\alpha_2 - 2\alpha_1$ and phase factor controlled by $2\psi = 2\alpha_1 + 2\alpha_2$.

Table S1

| $\|\Omega\rangle$ | $P_\Omega P_e \|R\rangle = \tau\|\Omega\rangle$ | $\tau$ |
|---|---|---|
| $\|R^*\rangle$ | $\langle e\|R\rangle^2 \|R^*\rangle$ | $\frac{1}{2}\left(1+\sin(2\chi)\right)e^{2i\psi}$ |
| $\|L^*\rangle$ | $\langle e\|L\rangle\langle e\|R\rangle\|L^*\rangle$ | $\frac{1}{2}\cos(2\chi)$ |
| $\|R\rangle$ | $\|\langle e_t\|R\rangle\|^2 \|R\rangle$ | $\frac{1}{2}\left(1-\sin(2\chi)\right)$ |
| $\|L\rangle$ | $\langle e_t\|L\rangle^* \langle e_t\|R\rangle\|L\rangle$ | $\cos(2\chi)e^{-2i\psi}$ |